\documentclass[twocolumn,pra,amsmath,showpacs]{revtex4}

\begin{document}
\title{Perfect Test of Entanglement for Two-level Systems}
\author{Sixia Yu$^1$, Jian-Wei Pan$^{1,2}$, Zeng-Bing Chen$^1$, and Yong-De Zhang$^1$}
\affiliation{$^1$Department of Modern Physics, University of
Science and Technology of China, Hefei 230027, P.R. China}
\author{ }
\affiliation{$^2$Institut f\"ur Experimentalphysik, Universit\"at
Wien, Boltzmanngasse 5, 1090 Wien, Austria}
\date{\today}

\begin{abstract}
A 3-setting Bell-type inequality enforced by the indeterminacy
relation of complementary local observables is proposed as an
experimental test of the 2-qubit entanglement. The proposed
inequality has an advantage of being a sufficient and necessary
criterion of the separability. Therefore any entangled 2-qubit
state cannot escape the detection by this kind of tests. It turns
out that the orientation of the local testing observables plays a
crucial role in our perfect detection of the entanglement.
\end{abstract}
\maketitle

The entanglement or the quantum correlation has become a key
concept in the nowadays quantum mechanics. From a fundamental
point of view the entangled states of two spacelike separated
quantum systems give rise to the question of the completeness of
quantum mechanics starting with Einstein-Podolsky-Rosen paper
\cite{epr} and culminating in Bell's theorem \cite{bell}. Form a
practical point of view the entanglement has found numerous
applications in the quantum information such as quantum
computation and quantum  teleportation \cite{qt,pan1}. A practical
question arises as to how we can detect the entanglement
experimentally.

As is well known, the entangled states of a bipartite system are
states that cannot be prepared locally. More precisely, entangled
states are {\it not} classically correlated states, i.e, separable
states which are convex combinations of product states. The
entanglement, though simply defined, is notoriously difficult to
detect from both the mathematical and physical point of view.
There are plenty separability criteria for the separability of
bipartite systems \cite{peres,horo1,san,wu}, among which the
Peres-Horodecki (PH) partial transpose criterion
\cite{peres,horo1} is an operational-friendly criterion and the
Bell inequality distinguishes itself as an experimentally doable
test for the entanglement.

Initially, Bell's inequalities and its generalizations aimed at
ruling out various kinds of local realistic theories
quantitatively, providing a sufficient and necessary condition for
the existence of local hidden variable model in the case of two
settings \cite{fine,werner1,bruk}. Since any separable state
admits a local hidden variable model it obeys the Bell inequality.
For all separable states of two qubits the
Bell-Clauser-Horne-Shimony-Holt inequality \cite{chsh}
 \begin{equation}\label{bi1}
 \langle A_1B_1+A_1B_2+A_2 B_1-A_2 B_2\rangle_\rho\le 2
 \end{equation}
holds true. Here $A_{i}=\vec a_i\cdot\vec \sigma$ and $B_i=\vec
b_i\cdot\vec\tau$ $(i=1,2)$ are two arbitrary sets of local
testing observables with $\vec\sigma$ and $\vec\tau$ being the
Pauli matrices for two qubits respectively; the norms of the real
vectors $\vec a_i$, $\vec b_i$ are less than or equal to 1;
$\langle{AB}\rangle_\rho={\rm Tr}(\rho AB)$ denotes the average of
the observable $AB$ in the state $\rho$ as usual.

Since the Bell inequality can be viewed as a property of separable
states, it provides a sufficient criterion for the entanglement.
One needs only choose properly the testing observables such that
the above inequality is violated in order to ensure an entangled
states. For multi particles the generalization of the Bell
inequalities can be employed to detect the totally separable
states \cite{mermin, gisin1} and fully entangled states
\cite{collins,svet,uffink}, and to classify the multiparticle
entanglement \cite{yu1}.

Though in case of pure states the Bell inequality is also
necessary for the entanglement \cite{gisin2}, there are entangled
mixed states that escape the Bell tests. For example there exists
Werner state that is entangled while admitting a hidden variable
model \cite{werner2}. Therefore the entanglement of these states
cannot be detected by the ordinary Bell tests. Bell's inequalities
with multi settings have also been investigated \cite{gisin3}
however a better criterion for the entanglement is not achieved.

The main reason why there are entangled states that escape the
ordinary Bell tests is that the entanglement is a {\it quantum
correlation} and Bell's inequalities do not require the quantum
nature of the local systems. To detect more effectively the
quantum correlations the quantum aspects of the local systems have
to be taken into account. In the present Letter we shall provide a
Bell-type inequality on the quantum correlations between two sets
of complementary local observables as a test for the entanglement,
which serves as a sufficient and necessary criterion of the
separability.

One distinguishing feature of quantum systems is the existence of
noncommuting observables together with corresponding uncertainty
relationships. To take into account this quantum nature of the
local systems we choose our testing observables to be
complementary observables, observables that are maximally
noncommuting. For a two-level system there are three mutually
complementary observables $A_i=\vec a_i\cdot\vec\sigma$ with three
normalized vectors $\vec a_i$ $(i=1,2,3)$ being {\it orthogonal}
to each other. In case of spin half system these observables
correspond to the measurements of three orthogonal spin
directions. They are complementary \cite{kraus} because if the
system is prepared at an eigenstate of one observable the results
of measuring any one of two other observables will turn out to be
completely random. In fact these three observables form a complete
set of complementary observables of the two-level system
\cite{cc,wootter0} in the sense that there is no other observable
that is complementary to all these three observables.

Now we shall introduce the notion of {\it orientation} for a
complete set of complementary observables which plays an essential
role in our to be proposed detection of the entanglement. For a
set of three mutually complementary observables $\{A_i\}$ we
denote $\mu_A=-iA_1A_2A_3$ as its orientation which can assume
only two values $\pm1$. If $\mu_A=1$ the orientation of the basis
formed by three real vectors $\vec a_i$ is right-handed, the same
orientation as that of three Pauli's matrices $\vec\sigma$. As a
result we have $A_1A_2=iA_3$ and so on. Obviously $A_i$ and $-A_i$
have different orientations. If $\{A_i\}$ runs over all set of
complementary observables with a certain orientation then
$\{-A_i\}$ runs over all set of complementary observables with a
different orientation. Similarly, we can introduce three mutually
complementary observables $B_i=\vec b_i\cdot\vec\tau$ $(i=1,2,3)$
and the corresponding orientation $\mu_B$ for the second qubit. In
the following we shall always denote $\{A_i,B_i\}_{i=1,2,3}$ as
two complete sets of complementary local observables. Now we are
ready to present our central result.

{\it Theorem: i) A two qubit state is separable if and only if the
following inequality holds for all set of observables
$\{A_i,B_i\}_{i=1,2,3}$ with the  same orientation:
\begin{equation}\label{myi}
\sqrt{\langle{A_1B_1+A_2B_2}\rangle_\rho^2+\langle{A_3+B_3}\rangle_\rho^2}
-\langle{A_3B_3}\rangle_\rho\le1.
\end{equation}
ii) For a given entangled state the maximal violation of the above
inequality is $1-4\lambda_{\min}$ with $\lambda_{\min}$ being the
minimal eigenvalue of the partial transpose of the density matrix.
The maximal possible violation for all states is $3$ which is
attainable by the maximal entangled states.}

To get some familiar with we observe at first that the inequality
above is essentially an indeterminacy relation. In fact if the
testing observables $\{A_i\}$ and $\{B_i\}$ have {\it different}
orientations, the following three observables $X_1=A_1B_1+A_2B_2$,
$X_2=A_3+B_3$, and $X_3=1+A_3B_3$ satisfy $X_1^2=X_2^2=X_3^2$. By
applying the Heisenberg-Robertson indeterminacy relation \cite{hr}
to observables $X_i$ $(i=1,2,3)$ we obtain $\det\Sigma\ge\det
\Gamma$ where $\Sigma$ is a $3\times3$  matrix with elements
$\Sigma_{ij}=\frac12\langle{X_iX_j+X_jX_i}\rangle_\rho-\langle{X_i}\rangle_\rho\langle{X_j}\rangle_\rho$
and $\Gamma$ is a $3\times3$  matrix with elements
$\Gamma_{ij}=\frac i2\langle{[X_i,X_j]}\rangle_\rho$
$(i,j=1,2,3)$. As a result we have
\begin{equation}\label{myunc}
\langle{A_1B_1+A_2B_2}\rangle_\rho^2+\langle{A_3+B_3}\rangle_\rho^2\le\langle{1+A_3B_3}\rangle_\rho^2.
\end{equation}
It should be emphasized here that this indeterminacy relationship
holds for {\it all} states, separable or entangled, so long as two
sets of complementary observables have {\it different}
orientations. If the testing observables are chosen to have the
same orientation the indeterminacy relationship above does not
hold generally except for the separable states.

Proof. i) If a state is separable then the partial transpose of
its density matrix remains a density matrix.  For convenience we
consider the local time reversal
$\tilde\rho=\tau_2\rho^{T_B}\tau_2$ instead, where $\rho^{T_B}$
denotes the partial transpose with respect to the second qubit and
$\tau_2$ is the second Pauli matrix for the second qubit. If the
state is separable then the time reversal $\tilde\rho$ is also a
density matrix and the indeterminacy relation (\ref{myunc}) holds
true for all complementary testing observables with {\it
different} orientations. For expectation values we have $\langle
A_iB_i\rangle_{\tilde\rho}=\langle{A_i\tilde B_i}\rangle_\rho$
$(i=1,2,3)$ and $\langle B_3\rangle_{\tilde\rho}=\langle{\tilde
B_3}\rangle_\rho$ where three observables $\tilde
B_i=\tau_2B_i^T\tau_2=-B_i$ $(i=1,2,3)$ form also a complete set
of complementary observables with a different orientation as
$\mu_B$, i.e., the same orientation as $\mu_A$. Therefore
indeterminacy relation (\ref{myunc}) holds true for testing
observables with the same orientation if the state is separable.
By noticing $\langle{1+A_3B_3}\rangle_\rho\ge0$ we therefore
conclude that the inequality (\ref{myi}) is necessary for the
separability from the indeterminacy relationship (\ref{myunc}).

If the inequality (\ref{myi}) holds true for all sets of
complementary observables with the same orientation then we have
to prove that the state is separable. Since we are dealing with
two-level systems it is suffice to demonstrate that the partial
transpose of the density matrix is positive~\cite{horo1}. That is,
we have to prove $\langle\Phi|\rho^{T_B}|\Phi\rangle\ge0$ or
equivalently $\langle\Phi|\tau_2\rho^{T_B}\tau_2|\Phi\rangle\ge0$
where $|\Phi\rangle$ is an arbitrary pure 2-qubit state.

According to the Schmidt decomposition, any normalized 2-qubit
pure state $|\Phi\rangle$ can be decomposed as
$s_1|+\rangle|-\rangle-s_2|-\rangle|+\rangle$ after a local
unitary transformation $U\otimes V$. Here the Schmidt numbers
$s_1, s_2$ satisfy $s_1^2+s_2^2=1$ and $s_1\ge s_2$ with
$C=2s_1s_2$ being exactly the concurrence \cite{wootter} of the
pure state and $|\pm\rangle$ are eigenstates of $\sigma_3$
$(\tau_3)$ corresponding to eigenvalues $\pm1$ respectively. If we
define two sets of complementary observables as
$A_i=U^\dagger\sigma_i U$ and $B_i=V^\dagger\tau_i V$, which have
the same orientation, then sixteen observables \{$1$, $A_i$,
$B_j$, $A_iB_j$\} form a base for all $4\times4$ matrix. By
expanding the density matrix $ |\Phi\rangle\langle\Phi|$ under
this base we obtain
\begin{eqnarray}\label{lemma}
1-4|\Phi\rangle\langle\Phi|=\mbox{
\quad\quad\quad\quad\quad\quad\quad
\quad\quad\quad\quad\quad\quad\quad\quad}\cr
\quad\sqrt{1-C^2}(A_3-B_3)-C(A_1B_1+A_2B_2)-A_3B_3.
\end{eqnarray}
Therefore a normalized state is in a one-to-one correspondence
with $\{A_i,B_i,C\}_{i=1,2,3}$ where $\{A_i\}$ and $\{B_i\}$ are
two set of local complementary observables with the same
orientation. We notice that the density matrix of a normalized
pure 2-qubit state has seven independent real parameters and a set
of three complementary observables, which is determined by an
orthonormal base of a three dimensional space, has three real
parameters. As a result we have
\begin{widetext}
\begin{eqnarray}
&&\langle\Phi|\tau_2\rho^{T_B}\tau_2|\Phi\rangle
=\mbox{Tr}\big(\rho\tau_{2}(|\Phi\rangle\langle\Phi|)^{T_B}\tau_{2}\big)\cr
&=&\frac14\left(1+\sqrt{1-C^2}\langle{A_3+B_3}\rangle_\rho+C\langle{A_1B_1+A_2B_2}\rangle_\rho+\langle{A_3B_3}\rangle_\rho\right)\cr
&\ge&\frac14\left(1+\langle{A_3B_3}\rangle_\rho-\left|\sqrt{1-C^2}\langle{A_3+B_3}\rangle_\rho+C\langle{A_1B_1+A_2B_2}\rangle_\rho\right|\right)\cr
&\ge&\frac14\left(1+\langle{A_3B_3}\rangle_\rho-\sqrt{\langle{A_3+B_3}\rangle_\rho^2+\langle{A_1B_1+A_2B_2}\rangle_\rho^2}\right)
\ge0.\label{prf}
\end{eqnarray}
\end{widetext}
Here the first inequality is due to inequality $-|x|\le x$ and the
second inequality is due to Cauchy's inequality and the last
inequality stems from inequality (\ref{myi}). Thus we have proved
that the inequality is sufficient for the separability.

ii) The smallest eigenvalue $\lambda_{\min}$ of the transposed
density matrix is given by
$\lambda_{\min}=\min_{|\Phi\rangle}\langle\Phi|
\rho^{T_B}|\Phi\rangle$ where the minimum is taken over all
normalized pure states. Since we have established the one-to-one
correspondence between a normalized state and local complementary
observables together with concurrence, the minimum can be taken in
two steps. At first we minimize the expectation value over the
concurrence, which is achieved in inequality (\ref{prf}), and then
over all testing local observables, which gives rise to the
maximal violation of inequality (\ref{myi}) as $1-4\lambda_{\min}$
for a given state. We notice that $\lambda_{\min}<0$ if the state
is entangled.

Since testing observables $\{A_i\}$ and $\{B_i\}$ have different
orientations we have $X_1^2+X_2^2=4$ remembering that
$X_1={A_1B_1+A_2B_2}$ and $X_2={A_3+B_3}$. As a result we obtain
$\langle{X_1}\rangle_\rho^2+\langle{X_2}\rangle_\rho^2\le 4$,
which is also satisfied when the orientations of the local testing
observables are different because of inequality (\ref{myunc}).
Together with $|\langle{A_3B_3}\rangle_\rho|\le1$ we observe that
$\lambda_{\min}\ge-\frac12$ from inequality (\ref{prf}). Thus we
conclude that the maximal violation of the inequality (\ref{myi})
is 3. Obviously this upper bound is attained by the maximally
entangled states. For example we choose two sets of complementary
observables according to $\vec a_i=\vec b_i$ $(i=1,2,3)$ for the
singlet state. \hfill Q.E.D.

Some comments are in order. Provided that the orientations of the
local testing observables are the same, all three observables
$X_i$ $(i=1,2,3)$ concerned in the inequality (\ref{myi}) are
commuting with each other, which can therefore be simultaneously
measured. If the local testing observables are chosen to have
different orientations, e.g., replacing $B_3$ by $-B_3$, then
there is no violation of inequality (\ref{myi}) at all. For a
known state the maximum violation $1-4\lambda_{\min}$ of the
inequality is always attainable. In fact one has only to choose
the testing observable to be those that correspond to the
eigenstate corresponding to the eigenvalue $\lambda_{\min}$ of the
partial transposed density matrix.

As is well known, the PH partial transpose criterion is necessary
and sufficient for the entanglement of 2-qubit state. Since all
physical processes must be represented by completely positive maps
\cite{kraus2} and the partial transpose is a positive map that is
not completely positive \cite{horo1}, in fact it is a difference
of two completely positive maps \cite{yu2}, it is impossible to
implement the partial transpose to a state through a physical
process. However in the experiments it is expectation values
involving both the state and observables that are concerned. What
is impossible for the operation of state can be implemented easily
to the observables. For an example, though the operation of local
time reversal to a state is impossible one can easily change the
orientation of the local testing observables to achieve the
effects of local time reversal.

Finally, by  choosing properly  the local testing observables and
noticing $\langle{A_3+B_3}\rangle_\rho^2\ge0$ we can obtain the
following simple test for the entanglement
\begin{equation}\label{ieq3}
|\langle{A_1B_1+A_2B_2+A_3B_3}\rangle_\rho|\le1
\end{equation}
which is necessary for a state to be separable. Again, if the
orientation of the testing observables are different then there is
no violation at all. The maximal violation of the inequality above
is 3 which is attained by the maximally entangled states. We
notice that this condition is also sufficient for the Werner
states $W=\alpha I+\beta|\psi_-\rangle\langle\psi_-|$ with
$|\psi_-\rangle$ being the singlet state. As is well known if
$\beta\le\frac13$ the state is separable. For $\beta>\frac13$ the
inequality (\ref{ieq3}) is violated by choosing local observables
$A_i$ and $B_i$ $(i=1,2,3)$ to point to identical directions.
Another distinguishing property of inequality (\ref{ieq3}) is that
all observables involved are commuting with each other and
therefore can be simultaneously measured.

In conclusion we have proved a Bell-type inequality as a
sufficient and necessary condition for the entanglement of 2-qubit
states. The difference between the inequality proposed here and
usual Bell inequality is that we involve the quantum behaviors of
the local systems by choosing the testing observables to be
complementary. If the quantum nature of the local systems is taken
into account the entanglement and quantum nonlocality can be
demonstrated to be equivalent \cite{chen}. Therefore our test
proposed for the entanglement here is also a test for the quantum
nonlocality. Though it remains hard to confirm the separability of
an unknown state in practice because one has to check the
inequality (\ref{myi}) for all sets of local complementary
observables, our {\it Theorem} does ensure in principle a
violation of the inequality (\ref{myi}) for any entangled state by
choosing properly the local testing observables. In a word, an
experimental realization of the Peres-Horodeki criterion of the
separability by changing the orientation of the testing
observables is provided for two-level systems. It is of interest
to consider a similar experimental realization of the
Peres-Horodecki criterion in case of multi-level or multi-particle
systems.

This work was supported by the National Natural Science Foundation
of China, the Chinese Academy of Sciences and the National
Fundamental Research Program (under Grant No. 2001CB309300).

\end{document}